\documentclass{aa}
\usepackage{amssymb}
\usepackage{amsmath}
\usepackage{xspace}
\usepackage{graphicx}

\usepackage{natbib}

\def\gr{$\gamma$-ray}
\def\fermi{Fermi\xspace}
\def\gc{Galactic centre\xspace}

\begin{document}

\title{Leptonic origin of the 100~MeV gamma-ray emission from the Galactic centre}
\author{D. Malyshev $^{1}$, M.Chernyakova $^{2,3}$,  A. Neronov $^{1}$, R. Walter $^1$}
\institute{
1. ISDC, Astronomy Department, University of Geneva, Ch. d'Ecogia 16, 1290, Versoix, Switzerland \\
2. Dublin City University, Dublin 9, Ireland\\
3. Dublin Institute for advanced studies, 31 Fitzwilliam Place, Dublin 2, Ireland\\
}

\abstract
{The Galactic centre is a bright \gr\ source with the GeV-TeV band spectrum composed of two distinct components in the 1-10 GeV and 1-10~TeV energy ranges. The nature of these two components is not clearly understood. }
{ We investigate the $\gamma$-ray properties of the Galactic centre to clarify the origin of the observed emission. }
{We report imaging, spectral, and timing analysis of   data from 74 months of observations of the Galactic centre by \fermi/LAT \gr\ telescope complemented by  sub-MeV data from  approximately ten years of INTEGRAL/ PICsIT observations.} 
{We find that  the Galactic centre is spatially consistent with the point source in the GeV band. The tightest $3\sigma$ upper limit on its radius is $0.13^\circ$ in the $10-300$~GeV energy band. The spectrum of the source in the 100~MeV energy range does not have a characteristic turnover that would point to the pion decay origin of the signal. Instead, the source spectrum is consistent with a model of inverse Compton scattering  by high-energy electrons. In this a model, the GeV bump in the spectrum originates from an episode of  injection of high-energy particles, which happened $\sim300$~years ago. This injection episode coincides with the known activity episode of the Galactic centre region, previously identified using X-ray observations.  The hadronic model of source activity could be still compatible with the data if bremsstrahlung emission from high-energy electrons was present in addition to  pion decay emission.}
{} 


\authorrunning{D.Malyshev et.al.} 
\titlerunning{Leptonic origin of gamma-ray emission from the GC}

\maketitle
\section{Introduction}
\begin{figure*}
\includegraphics[width=\textwidth]{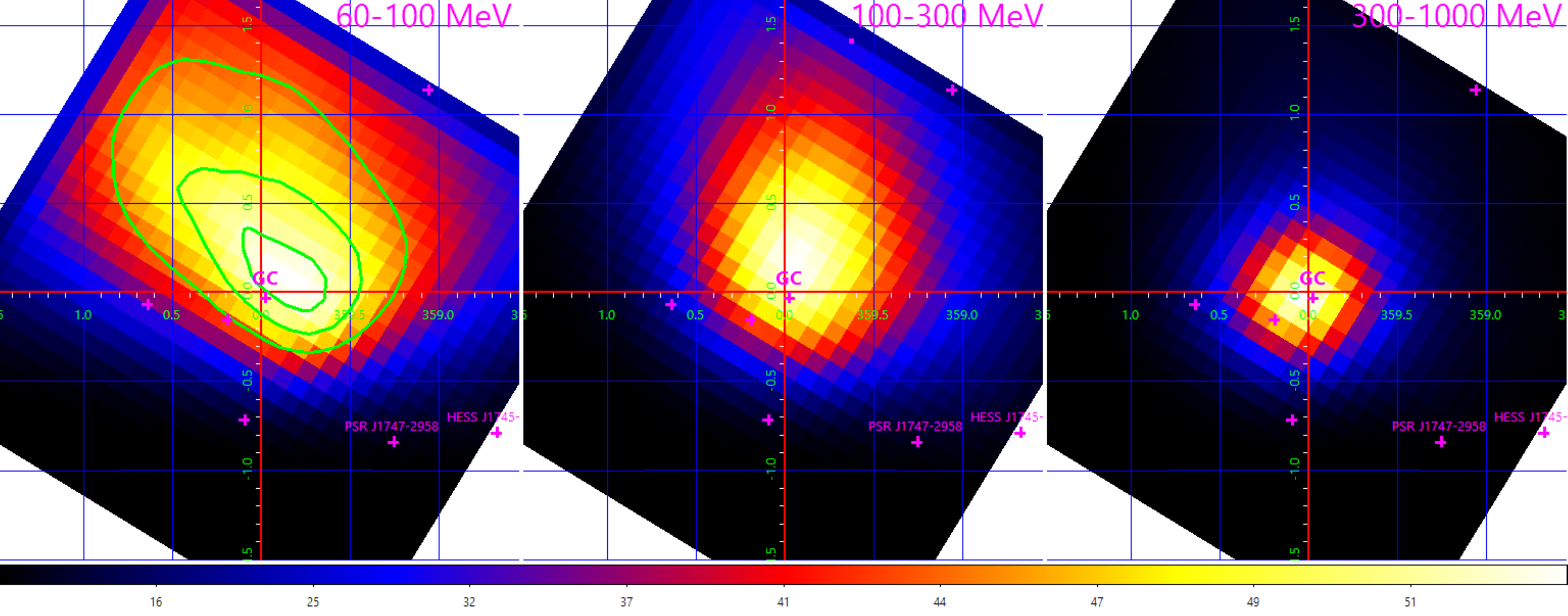}
\caption{Test-statistic (TS) maps in 60-100~MeV, 100-300~MeV, and 300~MeV-1~GeV energy bands. Only diffuse and point-like sources beyond $0.7^\circ$ from the position of GC were subtracted. The maximum of TS value corresponds to the most probable position of added point-like source.
The contours for the localisation significance are shown with  green solid lines for 60-100~MeV map and correspond to significances of 1, 2, and $3\sigma$. The 2FGL catalogue coordinates of the GC only deviate from the maximum TS position  at $\sim1\sigma$ level.
The position of maximum shifts with the increase of energy to the position of \gc.}
\label{fig:all_ts_maps}
\end{figure*}
The \gc (GC) is a unique astronomical source hosting the nearest supermassive black hole of  mass $M\simeq4\times 10^6M_\odot$ accreting matter in a radiatively inefficient regime and producing infrared, X-ray, and $\gamma$-ray emission with luminosity $L\simeq10^{36}$~erg/s \citep{genzel94,Mayer98,gc_hess,melia07}. 

The X-ray and infrared source could be associated with the supermassive black hole because of its variable nature \citep{X-ray_flares,infrared_flares}.  The source of the GeV and TeV $\gamma$-rays is not variable and it is not clear if the emission comes from the vicinity of the black hole, if it is produced in an extended region around the black hole, or, finally, if it comes from an unrelated source positionally coincident with the supermassive black hole in Sgr~A*. The best positional uncertainty of 6'' on  TeV source HESS J1745-303 \citep{acero10} still allows several sources to be responsible for the observed very high energy(VHE) emission:  the central black hole itself \citep{Aharonian05}, a plerion within several arcseconds from the GC \citep{wang06,hinton07}, a putative black hole plerion around Sgr~A* \citep{atoyan04}, and the central diffuse region around the GC \citep{Aharonian05a,ballantyne07,chernyakova:12}. 

In the GeV energy range, the GC source is localised with an accuracy of 72''  \citep{fermi2fgl} by the Large Area Telescope (LAT) on board the Fermi satellite  \citep{Atwood09}. An even larger range of possible counterparts could be discussed in this energy band. In addition to the astrophysical explanations, the observed GeV emission might also originate from the annihilation of dark matter particles \citep[see e.g.][]{hooper11,hooper11a,abazajan14}. In this scenario, the signal from the central ``spike'' of the Galactic dark matter halo is expected to be extended.

The analysis of 25 months of \fermi/LAT observations by \citet{chernyakova:12}  has revealed a GeV ``bump'' in the source spectrum. Assuming the identification of the \fermi source with the GC and HESS J1745-303, \citet{chernyakova:12} proposed a model that allowed the authors to explain the combined \fermi/LAT -- HESS spectrum with a model of $\pi^0$ decay $\gamma$-ray emission from protons diffusing away from the supermassive black hole. A similar model was considered by \citet{fatuzzo12}. An alternative explanation of the GeV-TeV spectrum with leptonic models was proposed by \citet{kusunose12}. Within this model, the GeV component of the spectrum is explained by the inverse Compton (IC) emission from electrons injected during NIR/X-ray flares of the central source, while the TeV emission originates from an unrelated source, e.g. the PWN G359.95-0.04. Still another  hybrid leptonic-hadronic model \citep{guo13} combines both approaches, explaining the TeV emission as caused by collisions of hadronic cosmic rays (CR) with the surrounding gas and attributing the GeV flux to IC emission of electrons accelerated near the GC.

In this paper, we report an updated study of the spectral, timing, and imaging characteristics of the GC source in the energy range  60~MeV -- 300~GeV, using six years of observations with \fermi/LAT.  We discuss  the implications of our findings for the theoretical models.  

\section{Data analysis and results}

\label{sec:Data_Analysis}

\subsection{Fermi/LAT data analysis}

We use 74~months of \fermi/LAT data (from August, 4th, 2008 to September, 25th, 2014).   We perform the binned likelihood analysis, using the most recent \fermi Science Tools Software v9r33p0 and the P7REP response functions. 

The analysis is based on the fitting of selected model of the observed sky region to the data.  We consider a region of radius $13^\circ$ around the GC and  \texttt{P7REP\_SOURCE} event class selection\footnote{http://fermi.gsfc.nasa.gov/ssc/data/analysis/documentation/
Cicerone/Cicerone\_LAT\_IRFs/IRF\_overview.html } for  analysis at  energies above $500$~MeV. 
The spatial model includes diffuse Galactic and extragalactic backgrounds and all sources from the two-year \fermi catalogue\footnote{When the paper was ready for submission, the four-year (3FGL) \fermi catalogue was released. In this catalogue, the emission is split into two sources located within $\lesssim 0.1^\circ$ central region.  We verified, using the 3FGL catalogue, that our analysis is valid for the total emission from this region.  } \citep[2FGL,][]{fermi2fgl}.

In the energy band below 500~MeV we use a broader, $30^\circ$, region and higher quality \texttt{P7REP\_CLEAN} class photons. The larger size of the analysis region and a more restricted event selection are necessary because of  a significant increase of the \fermi/LAT PSF (95\% containment for SOURCE class photons, which are $\sim3^\circ$ at 500~MeV and $\sim10^\circ$ at 100~MeV).
To avoid a significant increase in the computational time for  fitting  the data, as a result of the large number of catalogue sources in the selected region, we include only sources with a 100~MeV flux  higher than 10\% of the GC. This flux limit is comparable to the systematic uncertainties of the flux measurements\footnote{See http://fermi.gsfc.nasa.gov/ssc/data/analysis/LAT\_caveats.html}.  

\subsubsection{Fermi/LAT imaging analysis}

For  imaging analysis we  build test statistics (TS) maps in the energy bands 60-100~MeV, 100-300~MeV, and 300~MeV-1~GeV, see Fig.~\ref{fig:all_ts_maps}. 

To determine the best-fit position of the source, we remove  all sources beyond   $0.7^\circ$ from the GC position from the region
model. In the lowest (60-100~MeV) energy band, we find that the maximum of TS is shifted by $0.2^\circ$ from the GC. This deviation decreases with energy, and in the $300-1000$~MeV band the source position coincides with the GC with an accuracy better than the pixel size ($0.05^\circ$). Note, however, that the observed deviation is not statistically significant, see Fig.~\ref{fig:all_ts_maps} left panel.  
The position of the GC lies just beyond the $1\sigma$ contour.


\begin{figure*}
\includegraphics[width=0.45\textwidth]{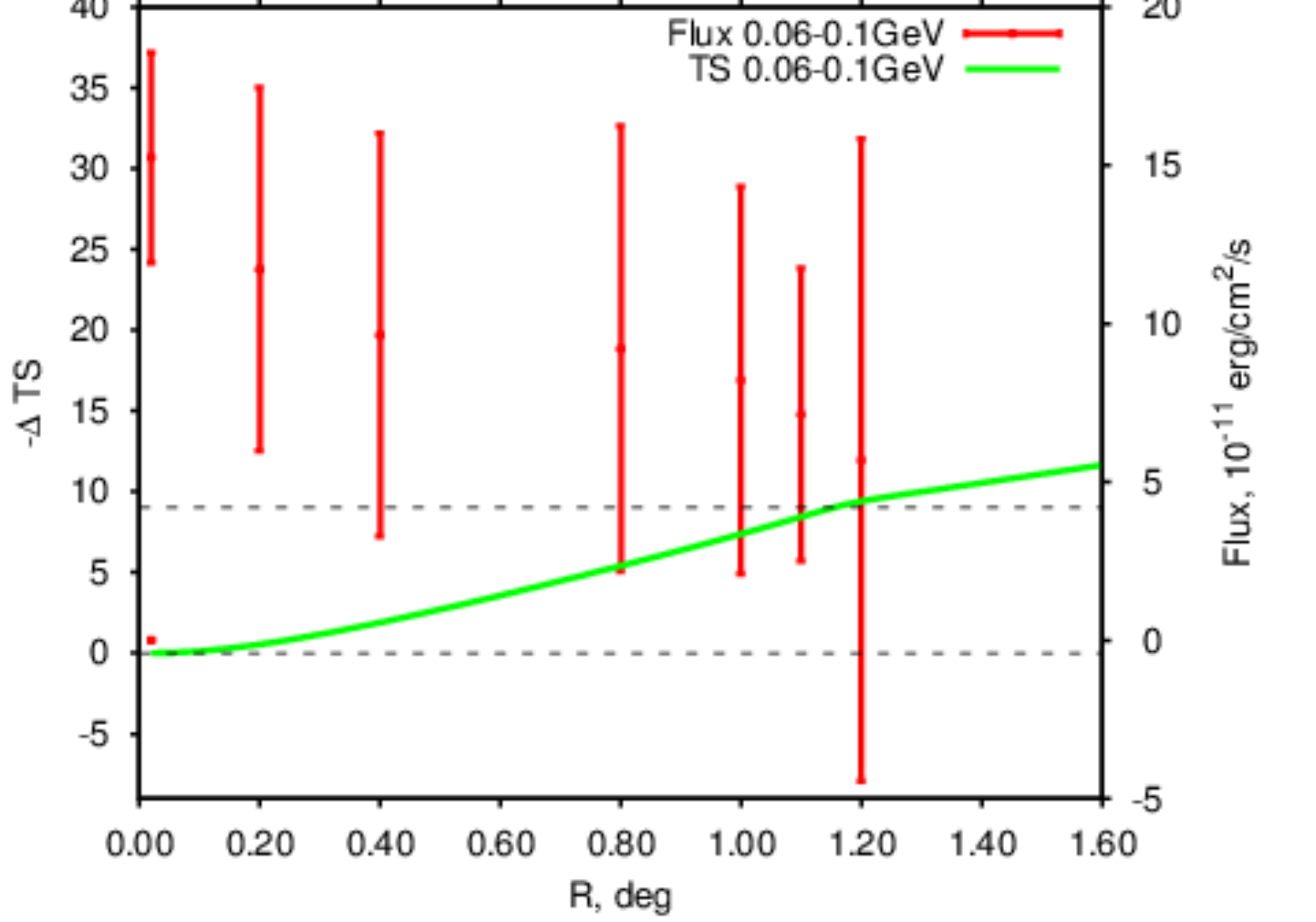}
\includegraphics[width=0.45\textwidth]{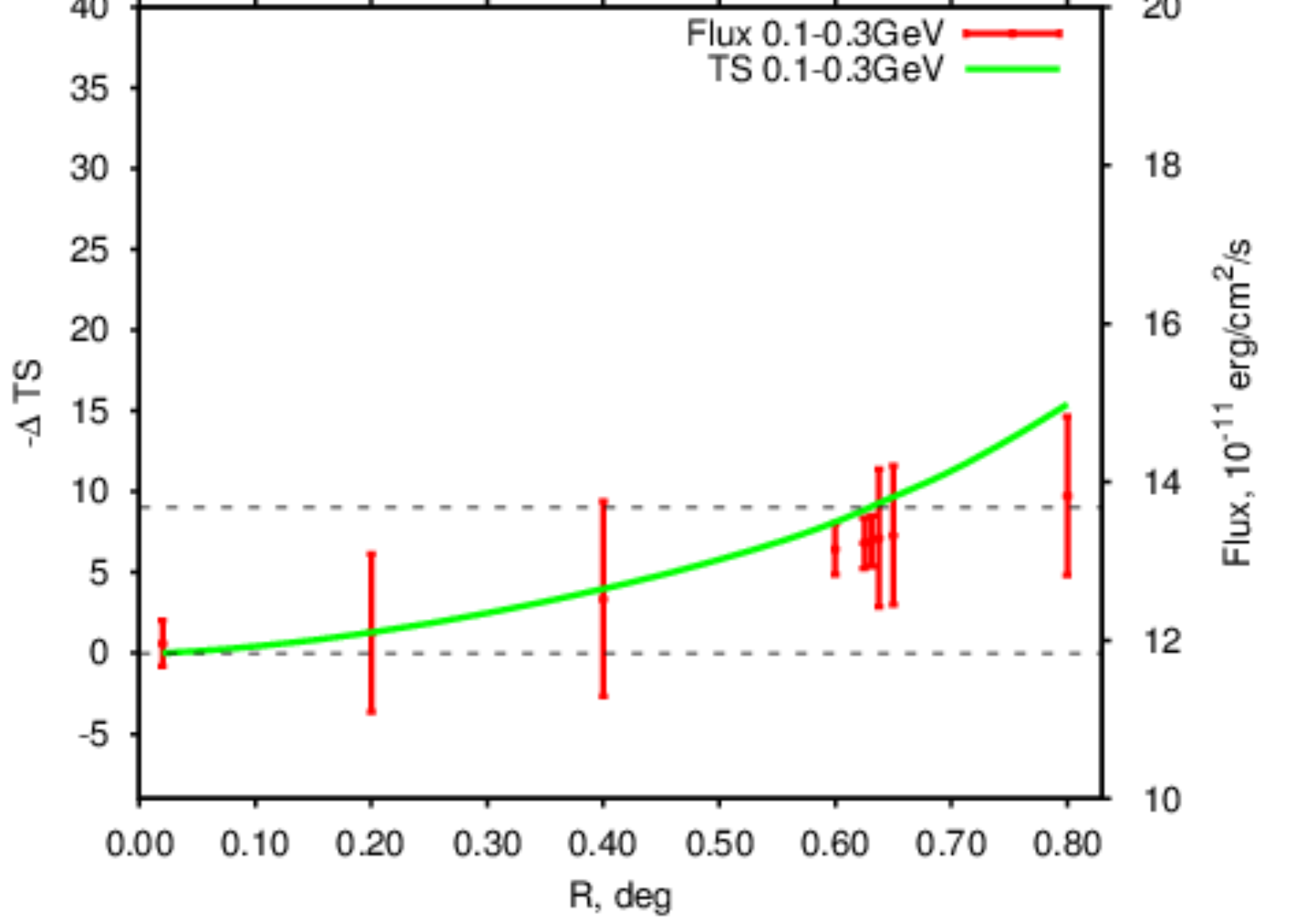} \\
\includegraphics[width=0.45\textwidth]{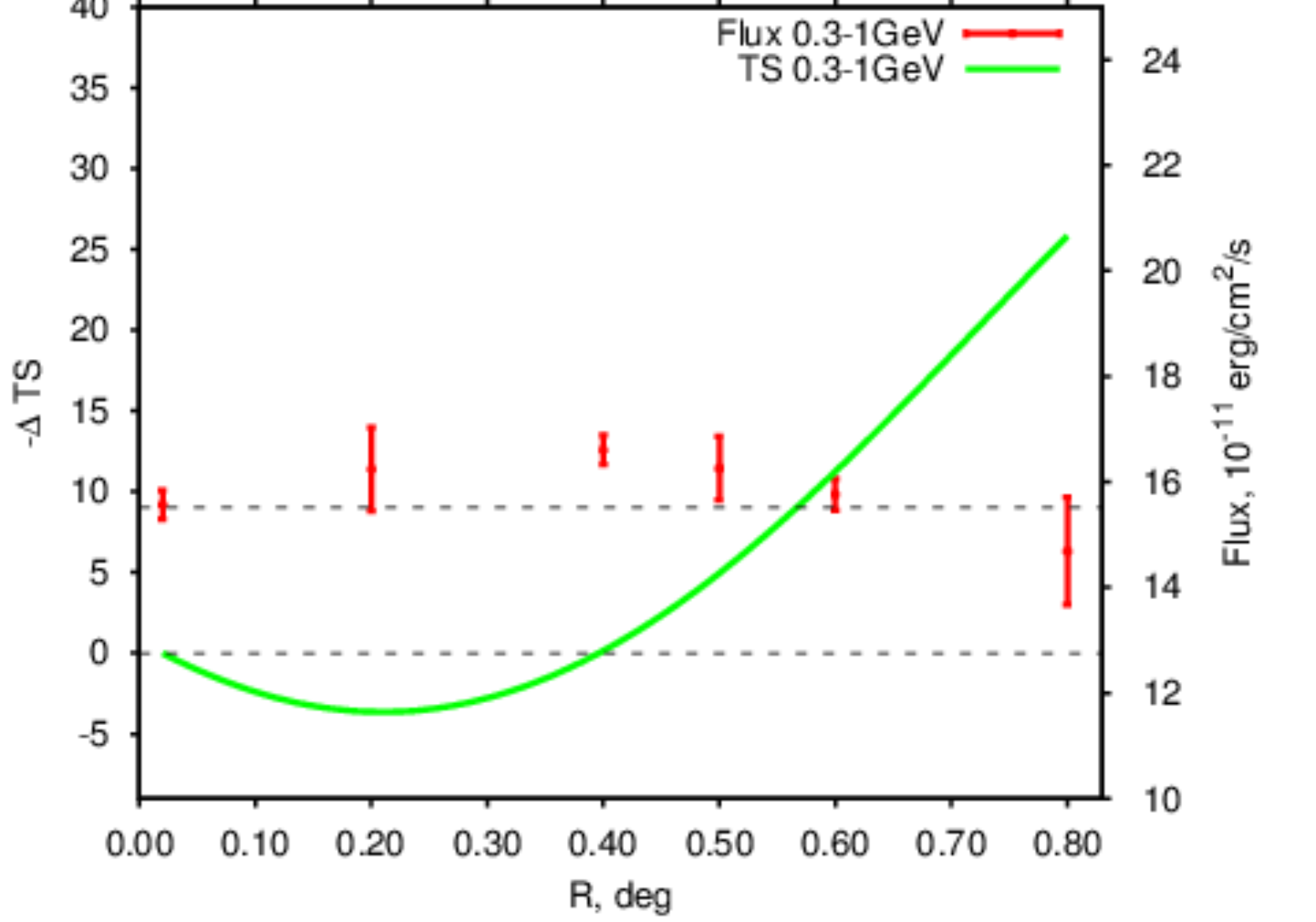} 
\includegraphics[width=0.45\textwidth]{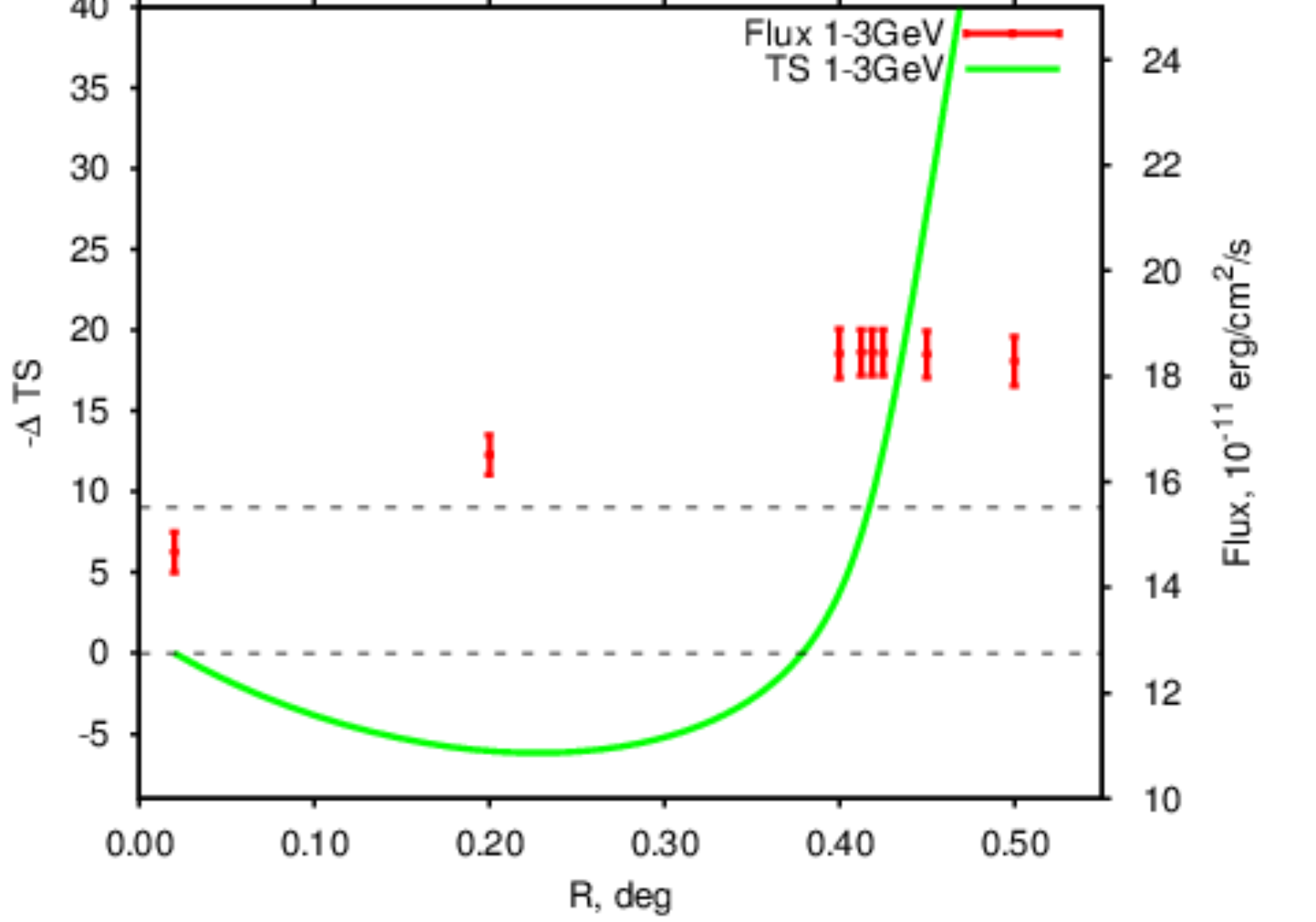}\\
\includegraphics[width=0.45\textwidth]{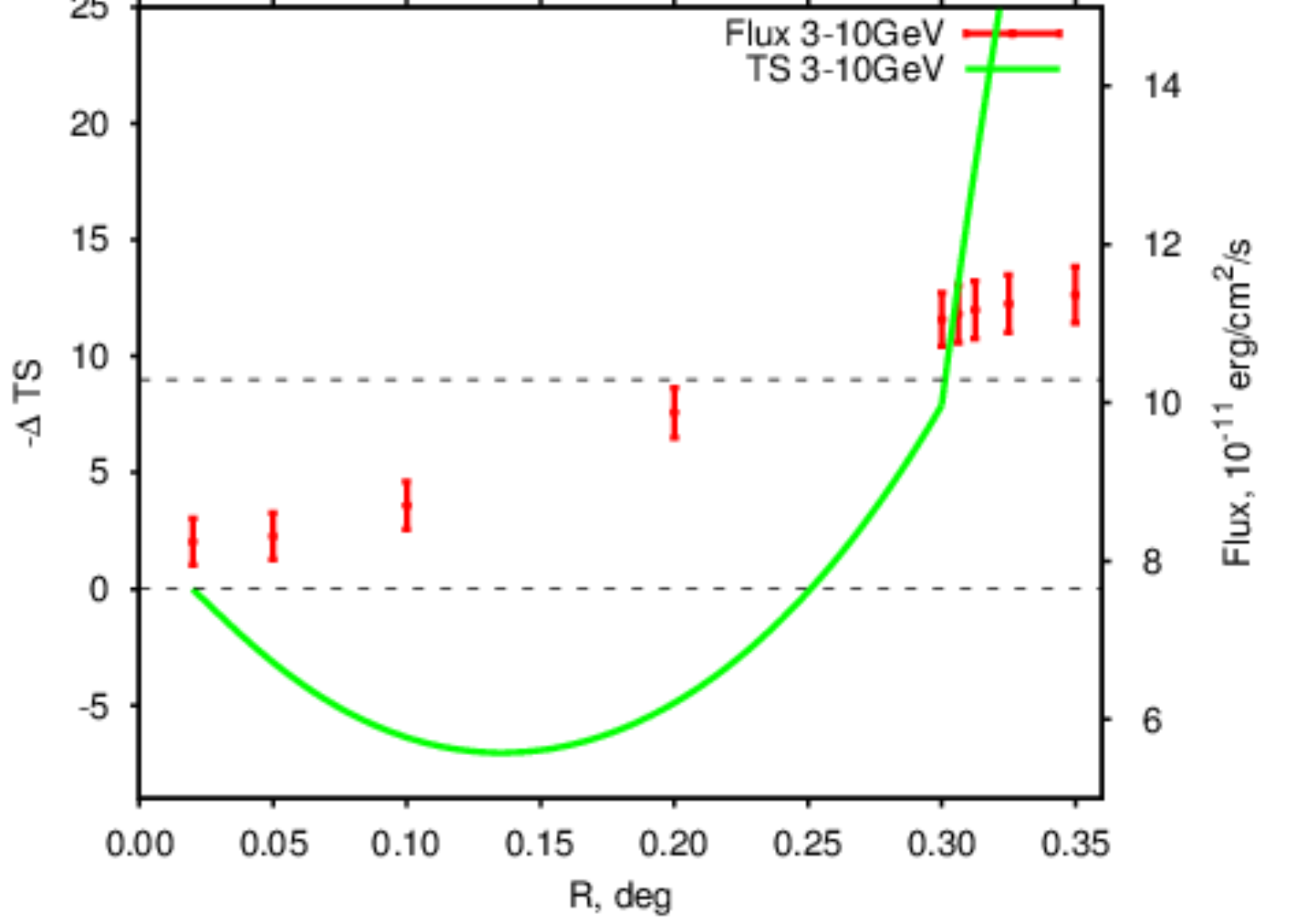} 
\includegraphics[width=0.45\textwidth]{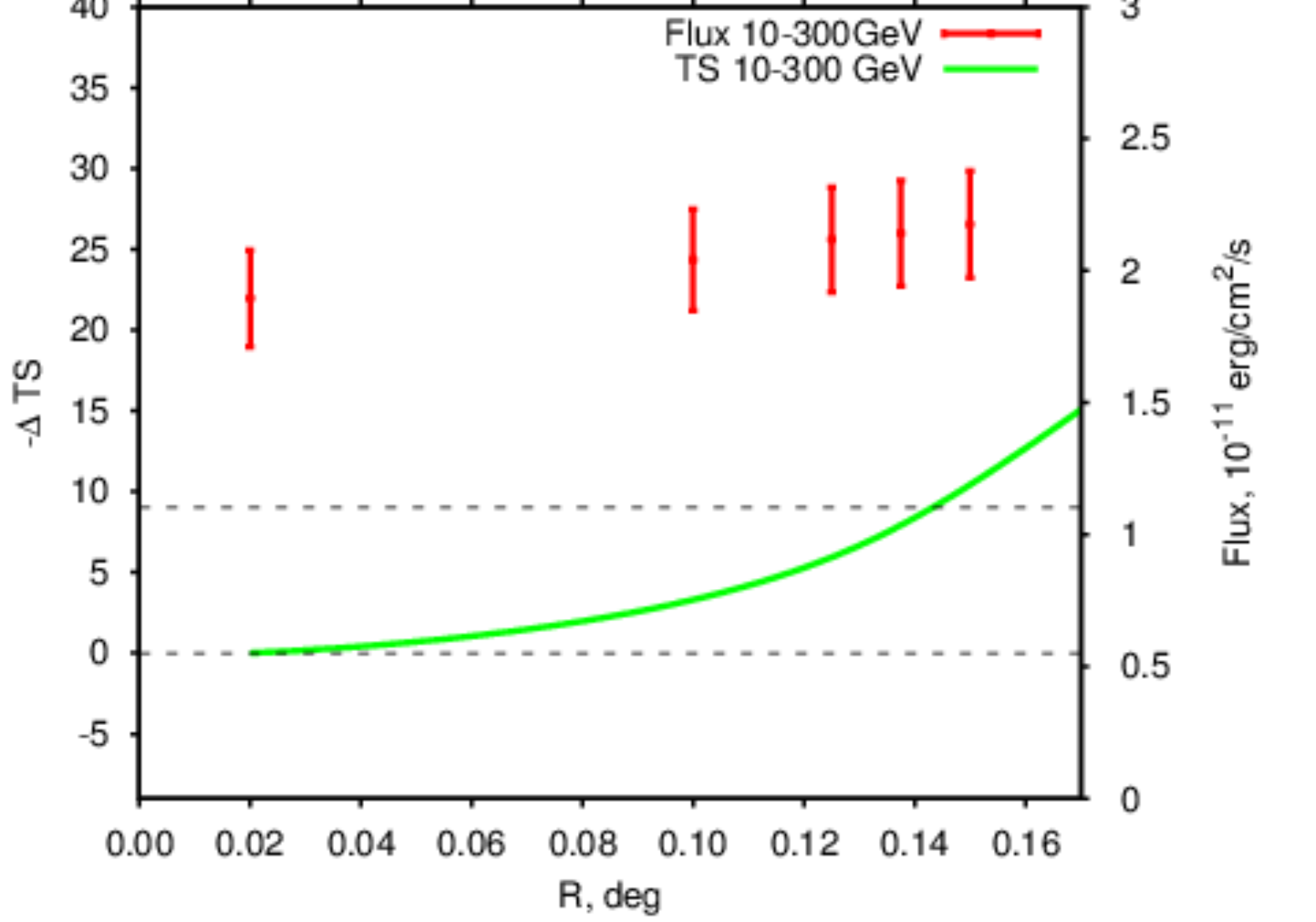} \\
\includegraphics[width=0.42\textwidth]{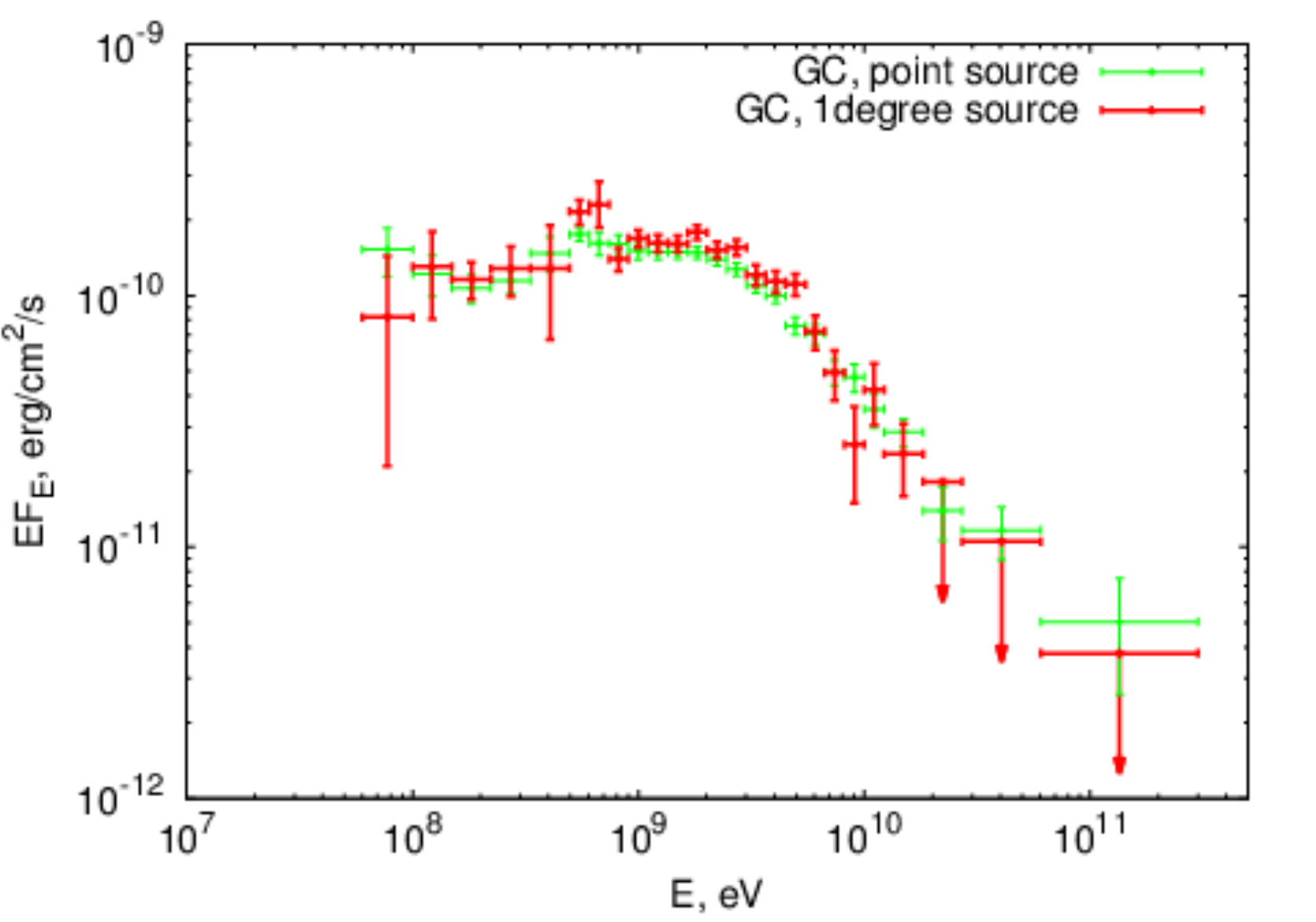}
\includegraphics[width=0.42\textwidth]{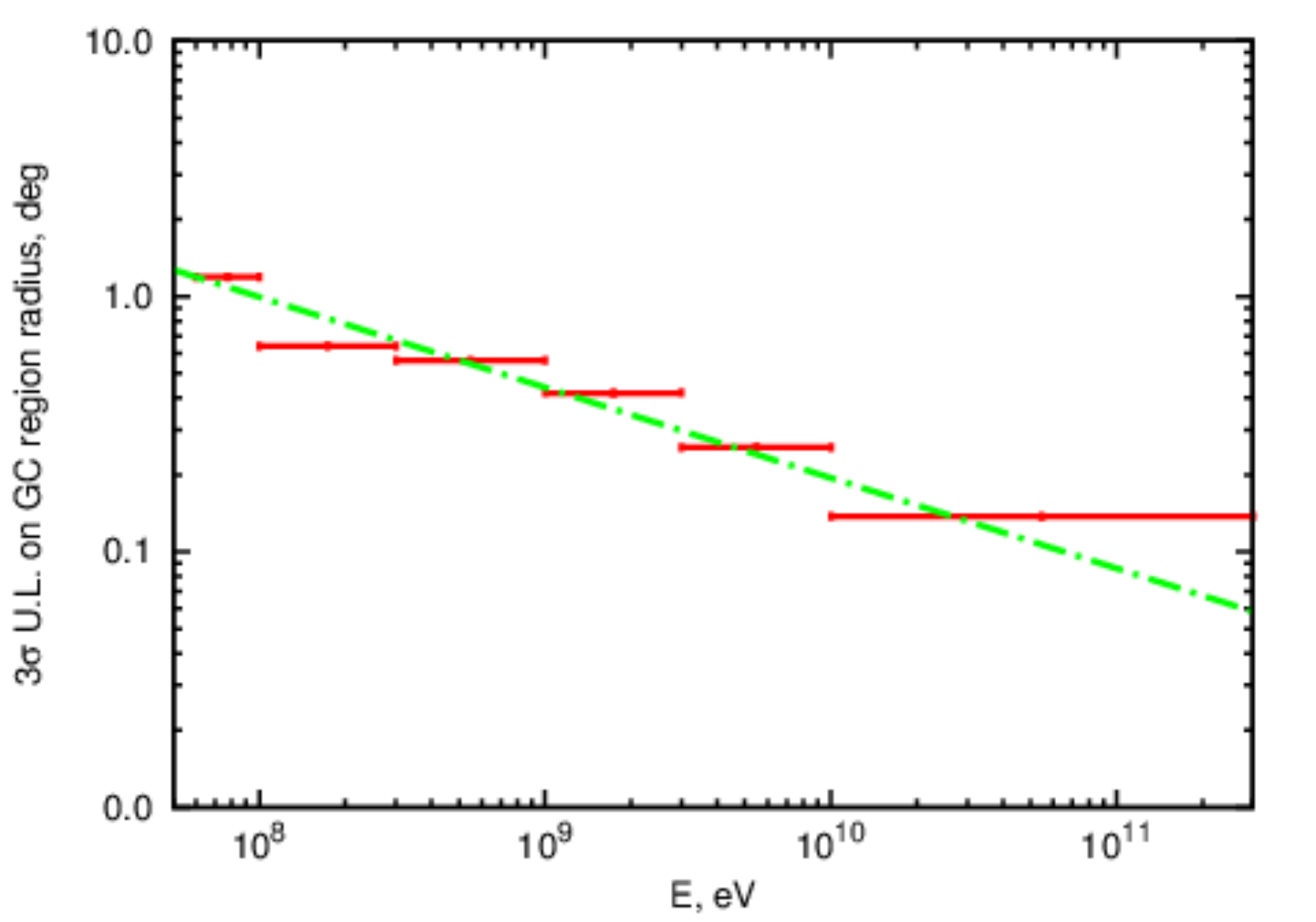}
\caption{The change (in the comparison with the point source) of the TS value attributed to the GC modelled as a disk of certain radius. The increase of $-\Delta TS$ corresponds to the overall worsening of the fit. The change of $-\Delta TS$ by 9 corresponds to the $3\sigma$ upper limit on the size of the disk. The flux, attributed to the disk is shown with red points with errorbars. \textit{Bottom left panel:} The spectra of $1^\circ$ radius template of the GC, and GC as a point source.
\textit{Bottom right panel:} $3\sigma$ upper limit on the size of the disk as a function of energy. The fit with the power law ($R = 0.44^\circ (E/ 1GeV)^{-0.35}$) is shown with green dot-dashed line. }
\label{fig:gc_extension}
\end{figure*}

At high energies ($\gtrsim 10$~GeV), the \fermi/LAT PSF is quite narrow ($\sim0.2^\circ$ 68\% PSF containment). We start with this energy band to put the tightest possible constraints on the GC spatial extent.
We replace the GC point-like source with a uniform disk. We first perform the fitting procedure for the model without the GC, and afterwards  add  the disk with variable radius, representing a diffuse GC source, to the model. The significance of the detection of the disk is given by the difference between log-like values of these models \citep[see e.g.][]{mattox96}. The following procedure is similar to that described in \cite{fermi2fgl} for the localisation of  catalogue sources.  
 \begin{figure*}
 \includegraphics[width=0.45\textwidth]{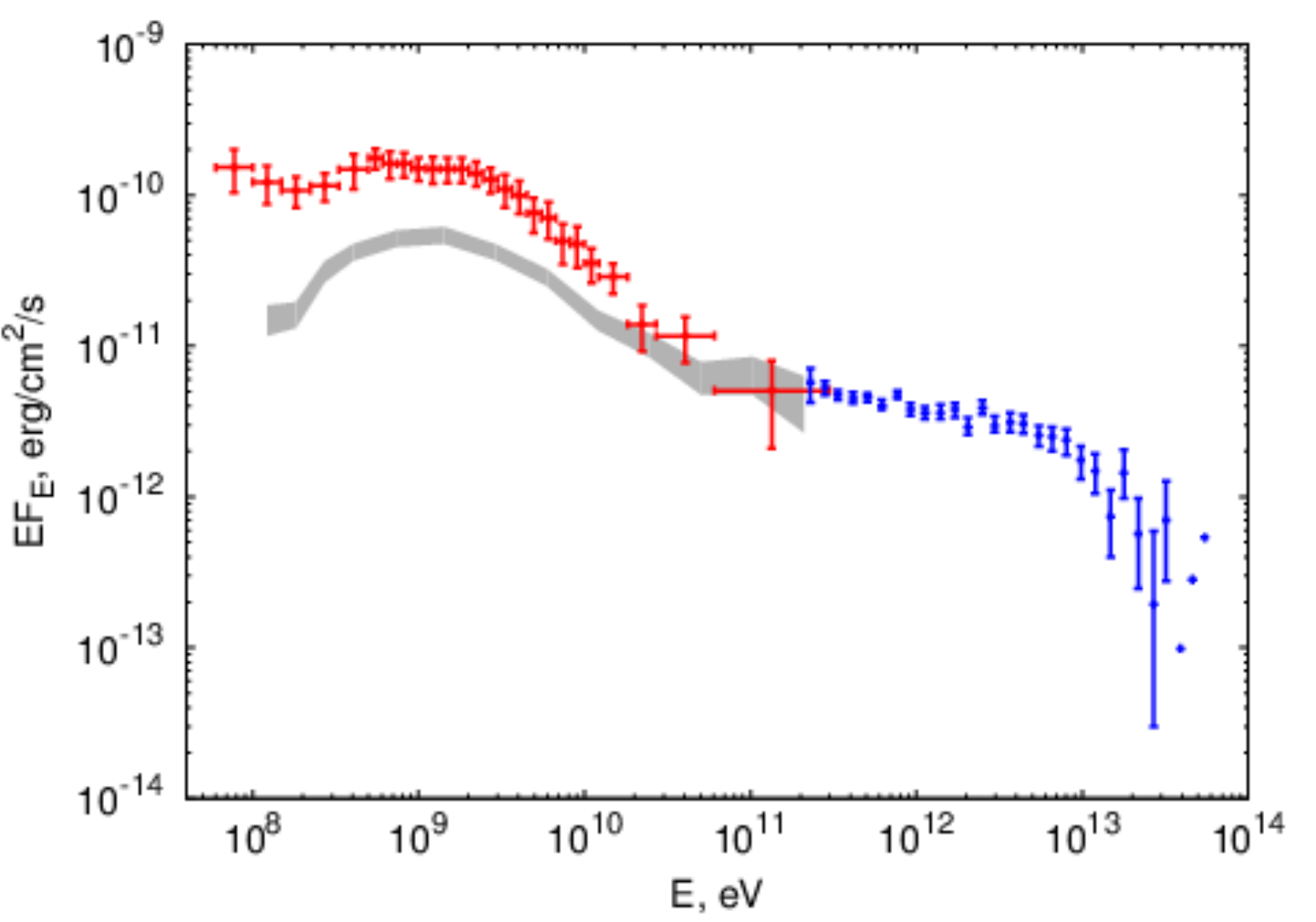}
 \includegraphics[width=0.45\textwidth]{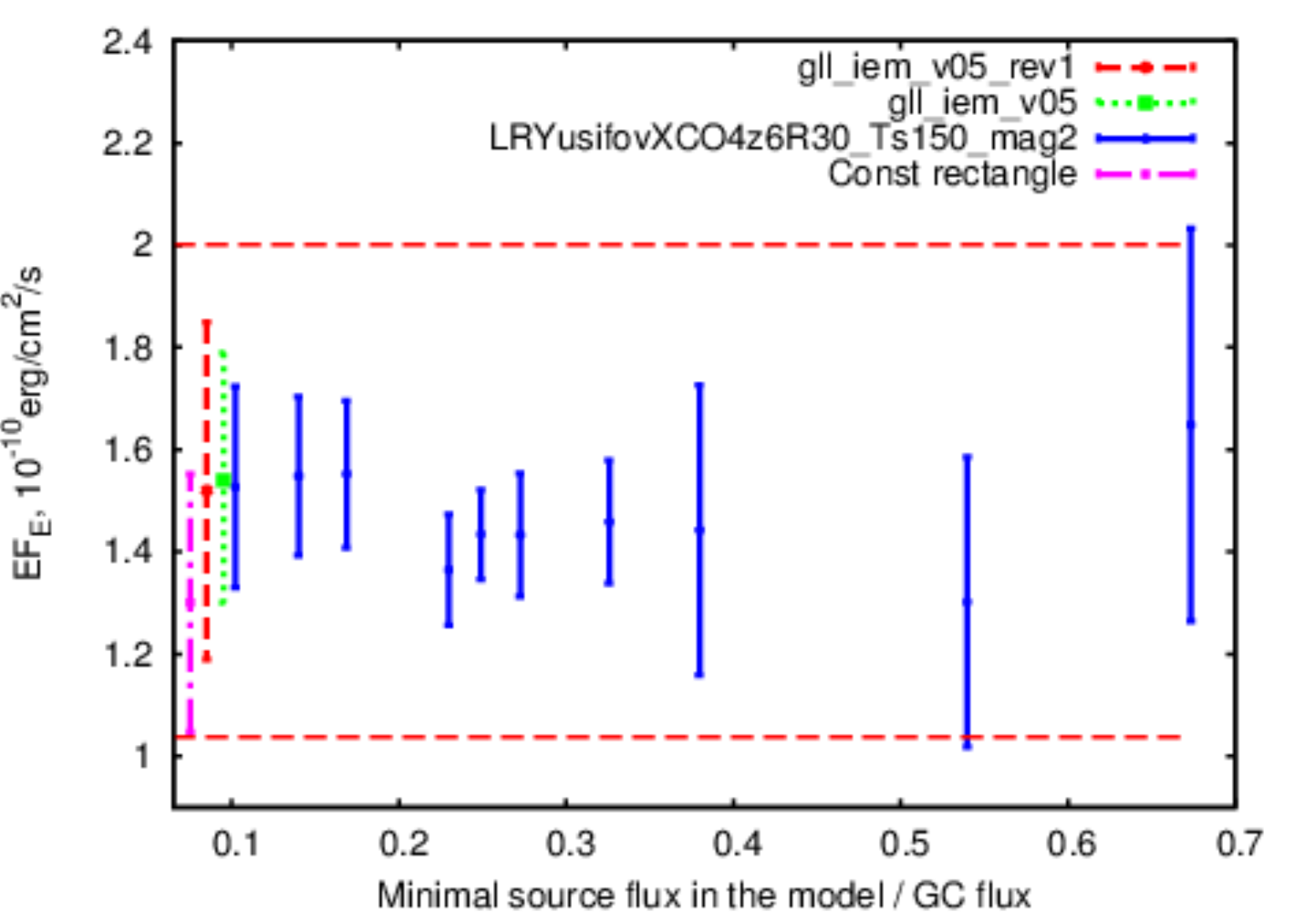}
 \caption{\textit{Left panel:} Combined \fermi/LAT -HESS spectrum of the \gc with added systematic errors. The spectrum of the Galactic bulge from ``constant rectangle'' model (renormalised to match the GC flux in the highest energy Fermi data point) is shown with grey shaded region, see text for the details.
 \textit{Right panel:} the flux in 60-100~MeV energy bin as a function of minimal catalogue flux of the sources included into the model(for \texttt{LRYusifovXCO4z6R30\_Ts150\_mag2} model of the galactic diffuse emission). The corresponding results for \texttt{gll\_iem\_v05\_rev1.fits}, \texttt{gll\_iem\_v05.fits} and ``constant rectangle'' diffuse models are shown with red dashed, green dotted, and magenta dot-dashed points. The horizontal dashed  lines indicate the flux errorbars in 60-100~MeV band from the left panel. }
 \label{fig:gc_fermi_hess_spec}
 \end{figure*}

Fig. \ref{fig:gc_extension} shows the results of the imaging analysis. In each energy band, the TS value remains roughly the same for disk radii $0<R<R_{max}$ with $R_{max}$, which shrinks with the energy, reaching $0.15^\circ$ in the 10-300~GeV energy band\footnote{This size is similar to the distance (0.093$^\circ$) between the GC and the closest to its source in the 3FGL catalogue. }, see corresponding panel of Fig.~\ref{fig:gc_extension}. The $R_{max}$ value is fixed by the requirement that the TS value decreases by 9. This source extension corresponds to the $3\sigma$ upper limit on the source size. 

The first six panels of Fig. \ref{fig:gc_extension} also show the dependence of the estimate of the source flux in each energy bin on the assumptions about the spatial extent of the source. One could see that the flux estimates vary by no more than $10\% - 20\%$ for the source extensions within $R_{max}$.

 The bottom right panel of  Fig~\ref{fig:gc_extension}, shows the $3\sigma$ upper limits on the source extension as a function of energy.
We find that the upper limits on the disk size become weaker with decreasing energy. This is obviously explained by the energy dependence of the PSF. The upper limit on the source size  can be described with the single power law $R_{max}=0.44^\circ (E/1GeV)^{-0.35}$ with the precision $\sim15\%$ at each point. 

In GeV range the fit prefers the diffuse source (of $\sim0.2^\circ$ radius) to the point source. This finding is in the agreement with e.g. \citet{hooper11,hooper11a,abazajan14}, who find that the data is better described by a slightly extended source. However,  evidence for the source extension is weak,  below the $3\sigma$ level.

\subsubsection{Fermi/LAT spectral analysis}

The spectral model of each source in the region of interest is chosen to be a power law with the slope $-2$ in each energy bin. The diffuse background model spectra are following the   templates, the \texttt{iso\_source\_v05\_rev1.txt} for the isotropic background and the \texttt{gll\_iem\_v05\_rev1.fits}  template for the Galactic diffuse emission background.
The analysis was performed with the \texttt{BinnedAnalysis} and \texttt{UpperLimits} python modules provided with Fermi Science Tools software. 
The resulting spectrum of the source is shown in Fig. \ref{fig:gc_fermi_hess_spec}.

In addition to statistical errors, we include a 10\% systematic error in energy range $100$~MeV -- $1$~TeV~\citep{fermi_syst}. This systematic error characterises uncertainties in the knowledge of the \fermi/LAT telescope. 
Taking  the complexity of the GC region into account,  we perform a separate study of additional systematic errors stemming from the uncertainties of the knowledge of  properties of the interstellar medium (and as a consequence, of the diffuse Galactic emission) and   the point source distribution in the  region of interest. 
We do this by repeating the analysis using several templates for Galactic diffuse emission and several sets of point sources. 
For the Galactic diffuse background, we consider \texttt{gll\_iem\_v05\_rev1.fits}, the most recent template, recommended \fermi/LAT collaboration for the analysis \texttt{gll\_iem\_v05.fits,} which is an earlier version of the template, and \texttt{LRYusifovXCO4z6R30\_Ts150\_mag2}\footnote{See e.g. http://fermi.gsfc.nasa.gov/ssc/data/access/lat/
Model\_details/FSSC\_model\_diffus\_reprocessed\_v12.pdf}, a template based on the prediction from GALPROP code~\citep{galprop}. As a  radical alternative to these models, we also consider a model in which we replace the background in the $3^\circ \times 1^\circ$ box around the GC in \texttt{gll\_iem\_v05\_rev1.fits} template with a rectangle of uniform surface brightness. We consider the flux of this rectangle to be an additional fitting parameter in each energy band.   We verify that all the four models  yield fluxes of the GC consistent within 10\%. Changing the limiting flux level for the catalogue sources, or changing the catalogue altogether from 2FGL to 3FGL also produces changes in the flux that are within the 10\% level; see right panel of Fig.~\ref{fig:gc_fermi_hess_spec}.

The broadening of the \fermi/LAT PSF from $\sim0.1^\circ$ at 100~GeV to $\sim10^\circ$ at 100~MeV together with the overall complexity of the region can lead to the source confusion problem. The spectral results in this case are given by the sum of the fluxes of several nearby point sources and/or the unaccounted diffuse emission. The variation of the number of model sources as well as the consideration of the different background models show that the flux attributed to the GC  varies within 10\% systematics in the $60-100$~MeV band. This reveals that the possible presence of  extended emission does not influence the measurement of the GC source spectrum.

Energy dependence of the Fermi/lAT PSF could lead to yet another problem for the spectral extraction if the GC source is extended. The region from which the source flux is collected might have larger spatial extent at low energies because of  wider PSF.  To estimate the influence of this effect on the measurement of the spectrum, we have forced the spectral extraction for an extended source of the radius $R=1^\circ$, which corresponds to $R_{max}$ at 100~MeV. This spectral extraction method ensures that the source flux is collected from the same region (rather than from within the PSF) at all energies. A comparison of the source spectrum extracted with this approach, with the spectrum obtained for the point source at the position of the GC, is shown in the bottom left panel of Fig. \ref{fig:gc_extension}. The two spectra are  consistent with each other. This shows that the energy dependence of the PSF does not have a strong influence on the measurement of the spectrum. 

\subsubsection{Fermi/LAT timing analysis}

\begin{figure}
\includegraphics[width=0.45\textwidth]{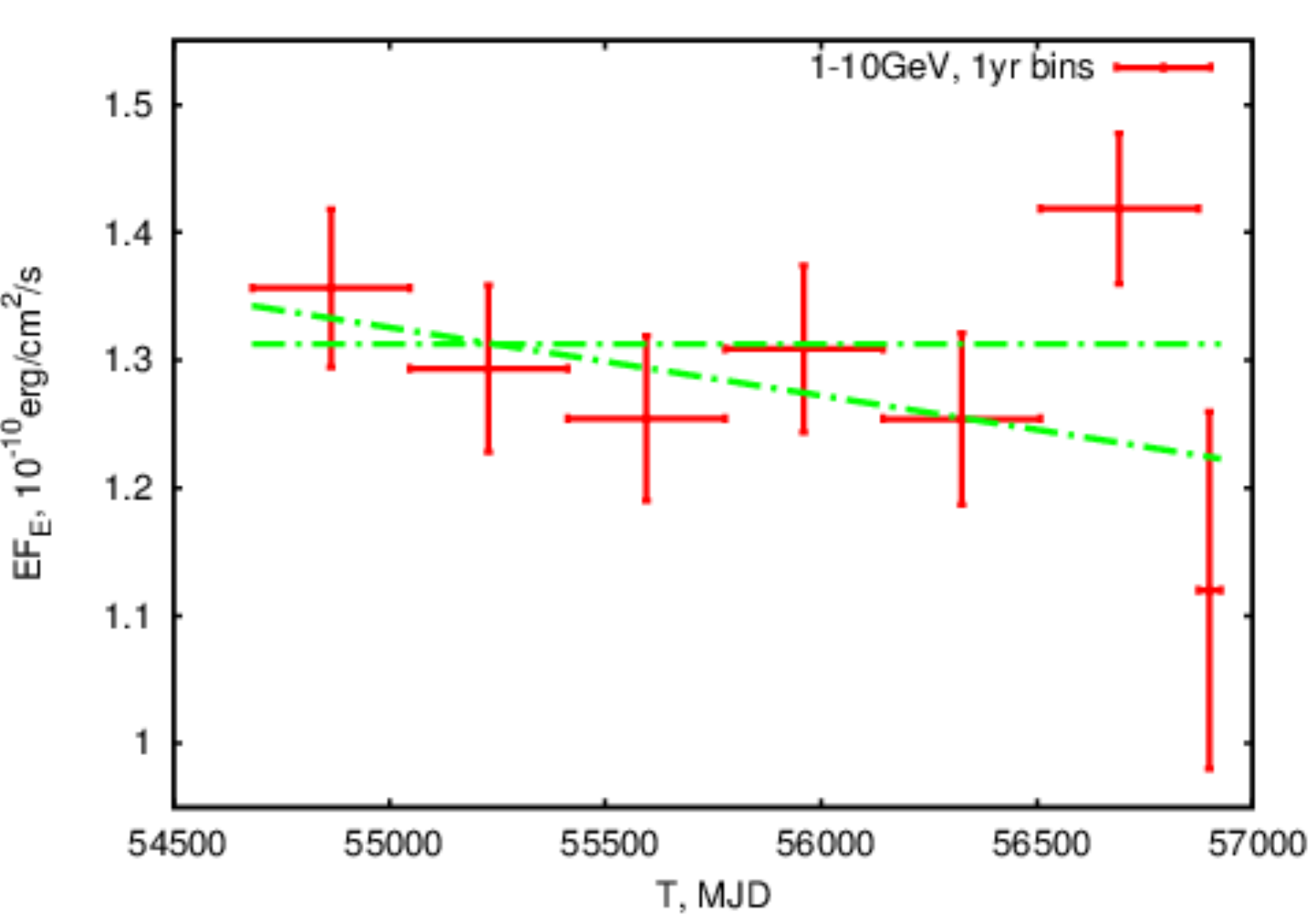}
\caption{The light curve of the \gc in 1-10~GeV energy band. Each point represents the bin with duration of 1~yr. The statistically allowed fits with constant and sloping lines are shown with dot-dashed curves.}
\label{fig:lc}
\end{figure}

The 1-10~GeV light curve, produced using  the likelihood analysis, is shown in Fig~\ref{fig:lc}. The source does not exhibit variability higher than $\sim10\%$ on yearlong timescales, the light curve is consistent with a constant flux model.

\subsection{INTEGRAL/ PICsIT data analysis}
We complement the GeV \fermi/LAT observations with  data from MeV-range PICsIT instrument.
 The Pixellated Imaging Caesium Iodide Telescope (PICsIT) is the high-energy detector of the IBIS telescope on board the INTEGRAL satellite, operating in energy range between 175~keV and 10~MeV \citep{dicocco03,labanti03}.

For the INTEGRAL/ PICsIT analysis, we selected all data obtained from February 2003 to August 2012. The data consists of detector images obtained in eight energy bands for single and multiple events. The resulting effective exposure times on the GC are 8.6 and 12.4 Msec for  single and multiple events, respectively. Detector uniformity and background maps have been built every ten spacecraft revolutions (three days each). 

The main sources of non-uniformities are two electronic effects relevant in the selection of single and multiple events on board. Multiple events are not detected at the boundaries between semi-modules and for the pixels located outside of the 4-by-4 pixel detector elements. The ratio between the single and multiple events is therefore not uniform across the detector. In addition, delays of time coincidence windows depend on individual front end electronic chips. Once corrected for these effects, the sky images are perfect but the instrument responses have to be corrected to take  the average energy dependence of the multiple event selection into account. As this effect cannot be calibrated properly, the responses were adjusted using spectra of the Crab nebula.

Mosaic images were finally built from the individual sky images. No source is detected at the \gc and upper limits were derived from the variance images and converted in flux units assuming a power-law spectral model.
Upper limits on the source flux derived from the INTEGRAL data are shown in Fig. \ref{fig:spectra_models}. 

\section{Discussion}

The origin of the multiwavelength emission from the \gc has been extensively discussed in the literature, starting from the pre-\fermi epoch, see e.g. \cite{Aharonian05,chernyakova:12,kusunose12}. The proposed models use different  types of radiation-producing particles, and the majority of them  can be divided into two broad classes, leptonic and hadronic. 

The spectrum of the GC shown in Fig.~\ref{fig:gc_fermi_hess_spec} is consistent with the previously reported spectrum  of \citet{chernyakova:12}, but extends to  lower energies (down to 60~MeV). This potentially opens a possibility of distinguishing between the leptonic and hadronic models because the hadronic spectral models are generically expected to possess a spectral feature in the 100~MeV energy range (energy comparable to the mass of the pions).

We obtained the spectrum of the GC  under the assumption that the diffuse background in the region is reasonably well described in one of the  templates above for the Galactic diffuse emission. 
These templates rely on the model of three-dimensional density distribution of the interstellar medium (except for the ``constant rectangle'' model template). The largest uncertainties of reconstruction of this three-dimensional distribution from different radial velocity data  is in the direction of the \gc. This uncertainty propagates to the uncertainty of the knowledge of the spatial morphology of the Galactic diffuse emission in the \gc region and further to the uncertainty of the measurement of the flux of the \gc source. In this respect, we are reassured by the fact that the measurement of the flux based on the ``constant rectangle'' model, which completely ignores the details of the spatial morphology of the diffuse emission around the source, agrees with the measurement based on the ``standard'' diffuse emission templates.

The uncertainty of the three-dimensional matter distribution in the direction of the \gc also leads to a potential problem of ``source confusion''. The non-variable source in the direction of the \gc might be confused with an unrelated molecular cloud, occasionally projected to the GC position. In this case, a part of the observed source flux should be attributed to the ``passive'' molecular cloud source rather than to a source actively accelerating particles in the \gc. The spectrum of this  passive source should approximately repeat the spectrum of diffuse emission from the interstellar medium (more precisely, of its pion decay and bremsstrahlung components). A test for this hypothesis could be a comparison of the spectral shapes of the \gc with the interstellar medium around the \gc.
Fig.~\ref{fig:gc_fermi_hess_spec} shows a comparison of the spectrum of the \gc with the spectrum of the rectangular box of the size $3^\circ \times 1^\circ$ (grey shaded region) from the ``constant rectangle'' diffuse background model. The spectrum of the box is normalised to match the flux of the GC at $\sim300$~GeV. One can see that the presented spectra are different. The spectrum of the GC in the energy band above 1~GeV is significantly softer than the spectrum of the diffuse emission right around the \gc. This indicates that the flux of the source is perhaps not dominated by the flux of a passive molecular cloud positionally coincident with the \gc on the sky.
\begin{figure}
\includegraphics[width=0.45\textwidth]{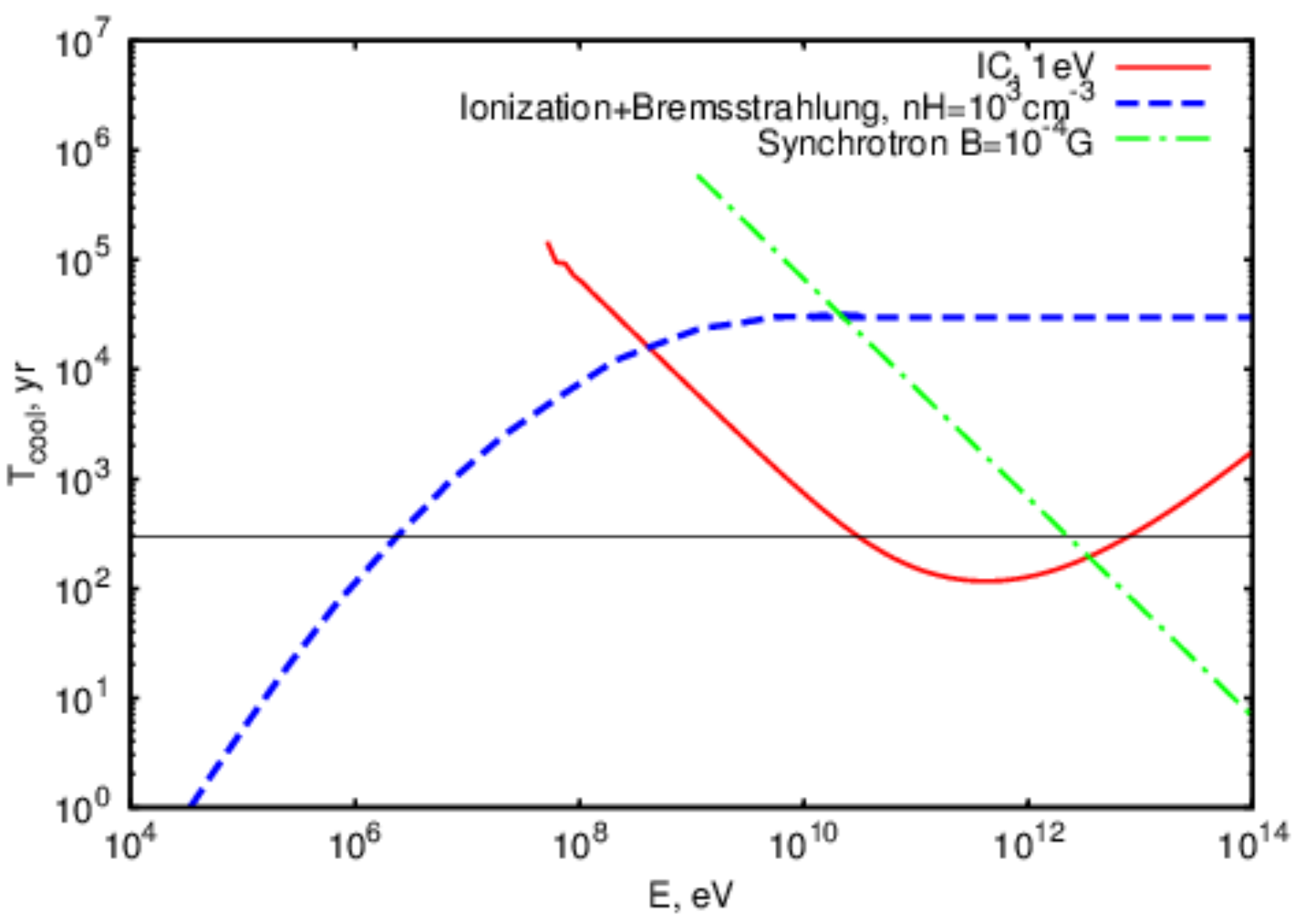}
\caption{Electron energy-loss cooling times for the medium parameters similar to those observed in the vicinity of the GC. The timescale of 300~yrs, corresponding to the flaring activity age, is shown with a thin solid line for  convenience.}
\label{fig:cooling_times}
\end{figure}
If a significant fraction of the source flux still comes from a molecular cloud situated in the GC region close or around Sgr A*, the mass of the cloud should be comparable to the entire mass of the nearby  Sgr B2 complex( $\sim10^7 M_{\bigodot}$ ), which is also a source of high-energy and very high-energy, gamma-ray emission with the flux comparable to that of the \gc at 100~GeV energy~\citep{hess_sgrb2}. Under this assumption, the upper limit on the size of the GeV source ($\lesssim 0.1^\circ$) translates to a lower limit of $\gtrsim 10^4$~cm$^{−3}$ on the density of the medium within a 10~pc scale region. This is somewhat higher than typical estimates of the density of the medium around Sgr A complex  \citep[see e.g.][]{tsuboi99,goldsmith90,genzel10}. 

Apart from the passive gamma-ray source produced by the interactions of interstellar cosmic rays with high-density medium near or around the \gc, an active acceleration process injects high-energy protons and/or electrons into the medium. Interactions of these particles lead to the gamma-ray emission of ``leptonic'' or ``hadronic'' origin. 

\subsection{Leptonic model}
The MeV-TeV spectrum of the GC can be interpreted with a simple one-zone leptonic model if one takes  the flaring nature of the source into account. In this case the emission is a combination of the emission from electrons injected during the strong flare, which occurred $\sim300$~yrs ago \citep[see e.g.][]{sunyaev93,koyama96,koyama08} and during  much weaker, recent activity. Electrons propagate through a medium with the density of about $10^3$~cm$^{-3}$, through the  soft photon field with density $\sim5\times 10^4$~eV/cm$^{-3}$ at $\sim0.3-3$~eV \citep{davidson92, mezger96, hinton07} and magnetic field of order 10-1000$\mu$G \citep{ferri09,eatough13}. The main energy-loss channels are ionisation, bremsstrahlung, inverse Compton (IC), and synchrotron mechanisms. Fig.~\ref{fig:cooling_times} shows a comparison of  the cooling times for these energy-loss channels, as a function of electron energy.

\begin{figure*}
\includegraphics[width=0.47\textwidth]{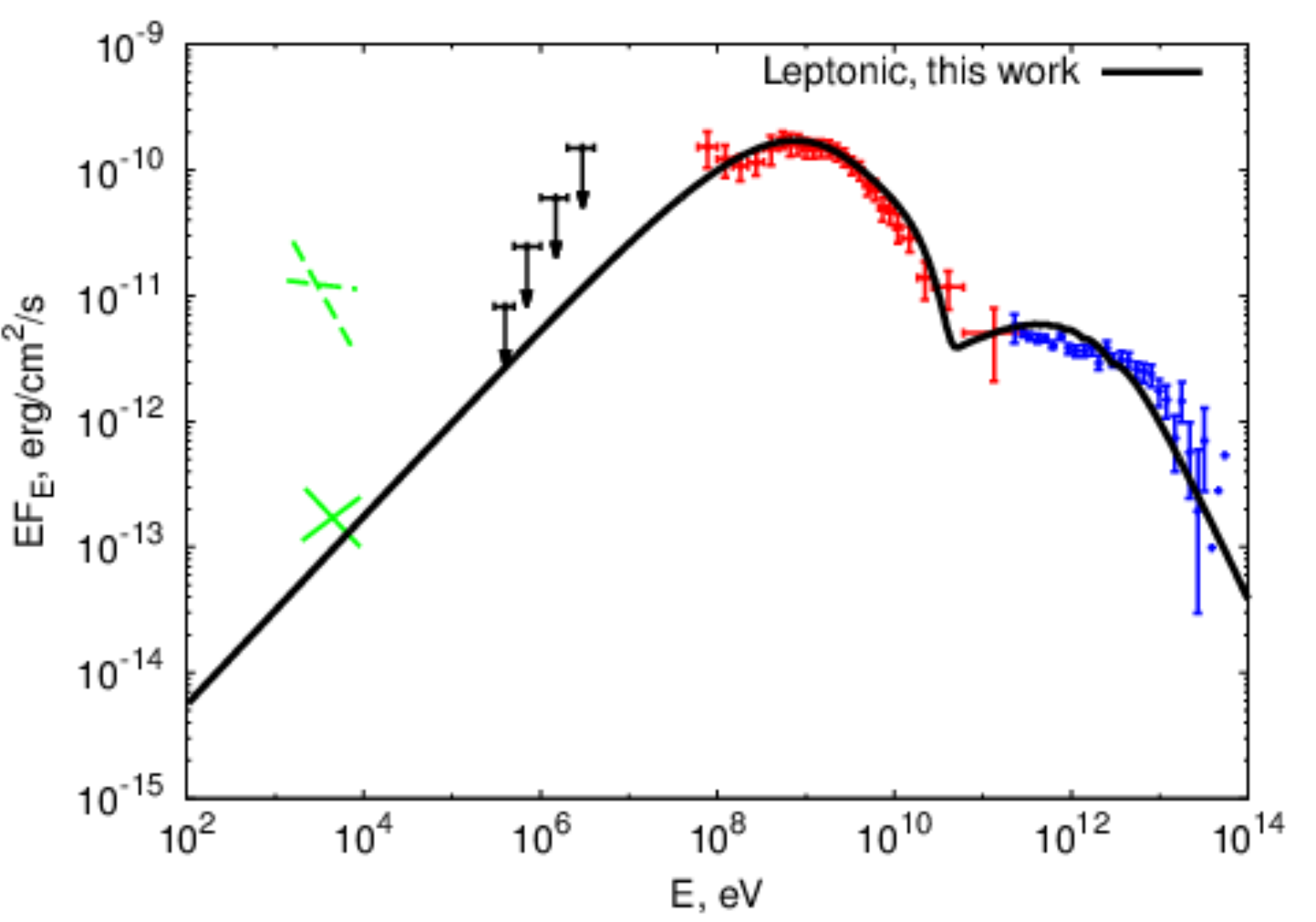}
\includegraphics[width=0.47\textwidth]{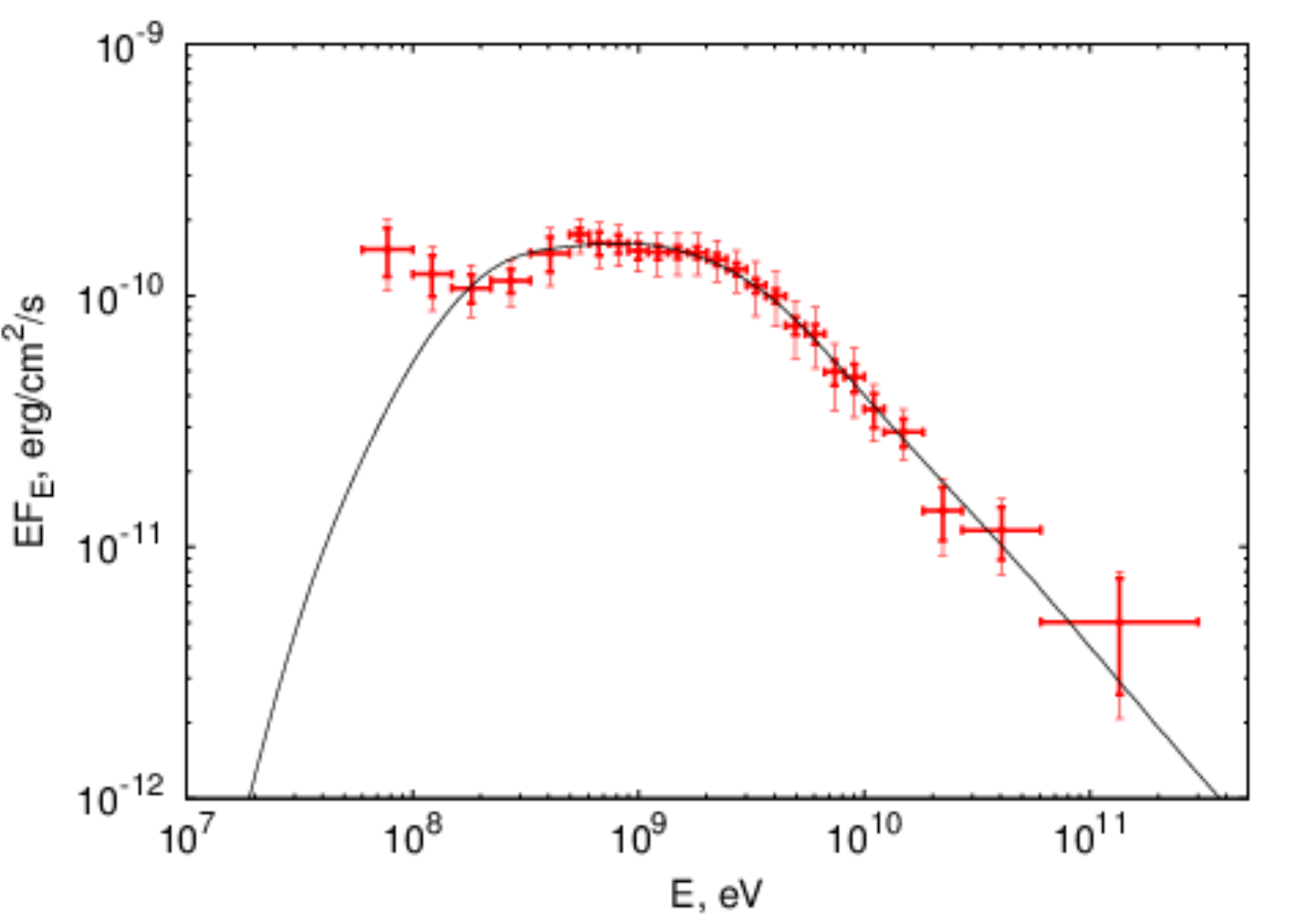}
\caption{\textit{Left panel:} X-ray to TeV energies spectrum of the \gc. The data are adopted from \citet{baganoff03} (green solid cross, X-ray quiescent), \citet{porquet08} (green dashed cross, X-ray flaring), this work (red points) and \citet{hess:GC_spectrum} (blue points, HESS). The INTEGRAL/ PICsIT upper limits (this work) are shown in black.
 The leptonic model proposed in this work for the low-density case is shown with the solid black line. 
\textit{Right panel:}  the black line shows the hadronic model of $\gamma$-ray radiation from the broken power-law distribution of protons. The model significantly under predicts the flux at $\lesssim 150$~MeV energies; see text for details. The statistical error is shown with thick errorbars, while the thin ones indicate systematic levels, which are shown in the left panel.  }
\label{fig:spectra_models}
\end{figure*}

We find that the observed source spectrum could be well reproduced by models with rather different choice of parameters. 
Depending on the ambient medium density, the observed spectrum can be explained either with a pure IC model (low medium density, see left panel of Fig.~\ref{fig:spectra_models}), or  by a combination of a IC scattering with a bremsstrahlung emission. 

In the left panel of Fig. \ref{fig:spectra_models}, we show an inverse Compton dominated model fit in   which electrons are scattered on photons with energy 0.5~eV and energy density $5\times 10^4$~eV/cm$^3$. The magnetic field is assumed to be $2.5\times 10^{-4}$~G. Electrons are injected with a spectral index -1.5 for 10~yrs during the flare that occurred 300~yrs ago. After the flare, the luminosity decreases by factor of $1.7\times 10^4$. During the flare the GC has reached a luminosity of $\sim2\cdot 10^{39}$~erg/s in the $\gamma$-rays (if the electron spectrum during the flare extended up to the 10~TeV energy band).

The  observed GeV bump in the GC spectrum is explained by IC scattering of the soft photons on the electrons injected  during the flare. The cooling time of electrons with energies about  $10$~GeV is comparable to the time since the flare. Suppression of the flux above the GeV energy is explained by the fact that electrons emitting at higher energies have already cooled down. The emission above 10~GeV is mostly due to the electrons injected during the ongoing  low activity period of the source. Above 10~TeV the emission is suppressed by synchrotron losses, which exceed the IC losses in the Klein-Nishina regime of IC scattering. To explain the TeV spectral shape in $\gamma$-rays, the magnetic field should be $\sim10^{-4}$~G.

The inverse Compton model of the GeV bump of the spectrum can be verified through the variability properties of the signal. If the GeV bump is really due to the electrons that were injected 300~years ago during a major flare of the source, gradual cooling of these electrons should lead to the displacement of the bump towards lower energies in the future. A noticeable displacement should  already occur on the 10-yr timescale.  The $1-10$~GeV flux is expected to decrease by $\sim5-10$\% on this timescale.  From the source light curve shown in Fig. \ref{fig:lc} one could see that the sensitivity of Fermi/LAT is marginally sufficient for the detection of this decrease.

\subsection{Hadronic model}

\citet{Aharonian05a} and \citet{chernyakova:12} developed a model in which the GC \fermi/HESS spectrum is explained by hadronic radiation from relativistic protons,  which diffuse by $\sim5$~pc away from the central source.  Data above 100~MeV are well described by the model for different sets of parameters (density profile of surrounding medium, characteristics of the diffusion coefficient, injection rate history). A generic feature of the hadronic models is a low-energy cut-off in the spectrum at 100~MeV at pion production threshold; see right panel of Fig~\ref{fig:spectra_models}. 

In this figure, we show the results of the fitting of \fermi/LAT data with the model incorporating broken power-law proton spectrum at \fermi energies, used as an approximation of~\citet{chernyakova:12} model. We find that this model provides a reasonably good fit of the data ($\chi^2/NDF$ = 11.0 /22, with included systematics, and $\chi^2/NDF$ = 32.2/22, without).  The two lowest-energy data points ($60-100$~MeV and $100-150$~MeV), where the cut-off in hadronic models is expected, deviate from the model by $2.4\sigma$ and $1.4\sigma$ ($3.5\sigma$ and $2.2\sigma,$ without systematic errors). In absolute values, the model flux in these energy bins is $\sim5$ and $\sim1.7$ times lower than the observed flux. 

The observed discrepancy with the data can be explained by the emission of several leptonic mechanisms, possibly operating in the region. The bremsstrahlung or IC emission from the electrons in the region can perfectly fill the gap in this energy interval.

\section{Conclusions}

We have presented the results of $\sim6$ years of \fermi/LAT observations of the \gc. We have shown that the spatial morphology of the GC source is consistent with the  point-like source. The $3\sigma$ upper limit on the radius of the source was found to be $\sim0.13^\circ$ at energies $\gtrsim 10$~GeV, and $\sim0.22^\circ$ at energies $3-10$~GeV. For lower energies, the radius of the source does not exceed $\sim0.7^\circ$; see Fig.~\ref{fig:gc_extension}.

The spectrum of the source  shown in the left panel of Fig.~\ref{fig:gc_fermi_hess_spec} covers a broader energy range compared to   the previous work by \citet{chernyakova:12}. In the overlapping range (0.1-300~GeV), these two spectra are consistent with each other.  
This indicates that the spectrum is independent of the version of the \fermi/LAT data analysis software as well as of the choice of the diffuse/point source backgrounds, confirming the correctness and robustness of previously obtained results.
The spectrum in the 60-300~MeV  energy band  does not show any evidence for low-energy cut-off, which is expected in pure hadronic models of the \gc emission. 

The MeV-TeV band spectrum of the GC is well modelled with a leptonic model in which the GeV bump in the spectrum is produced by the  IC emission from electrons injected during the strong flare from the GC, which happened $\sim300$~yrs ago. The model has a testable prediction of a decrease of the 1-10~GeV flux by $5-10\%$ on the $\sim10$~yrs timescale. The quality of the Fermi/LAT data is insufficient for testing the model prediction with a 6-yr data set (see Fig~\ref{fig:lc}). The prediction, however, can be tested within the next decade before the end of \fermi mission, or with one of its successors, e.g. ASTROGAM\footnote{http://astrogam.iaps.inaf.it/Doc/extract.pdf}, GAMMA-400~\citep{gam400}, HERD~\citep{herd}, PANGU~\citep{pangu}, or DAMPE\footnote{http://dpnc.unige.ch/dampe/index.html}.

Also, our analysis  does not rule out the models in which the leptonic GeV component of the GC spectrum has a different origin from the leptonic~\citep{kusunose12} or hadronic~\citep{guo13} TeV component.

The hadronic models \citep{Aharonian05a, chernyakova:12}  conflict with the data. The account of the electron bremsstrahlung or IC emission can increase the flux below $\sim100$~MeV and removes this conflict. We conclude that the presence of a leptonic component in the cosmic rays near the GC is unavoidably required in order to match the predictions of the models with the data.

\textbf{Acknowledgements.} The authors thank the International Space Science Institute (ISSI,
Bern) for support within the ISSI team ``Study of Gamma-ray Loud
Binary Systems'' and SFI/HEA Irish Centre for High-End Computing
(ICHEC) for the provision of computational facilities and
support. We also would like to thank our anonymous referee, whose report led to  significant improvement of the paper.

\def\aj{AJ}%
\def\actaa{Acta Astron.}%
\def\araa{ARA\&A}%
\def\apj{ApJ}%
\def\apjl{ApJ}%
\def\apjs{ApJS}%
\def\ao{Appl.~Opt.}%
\def\apss{Ap\&SS}%
\def\aap{A\&A}%
\def\aapr{A\&A~Rev.}%
\def\aaps{A\&AS}%
\def\azh{AZh}%
\def\baas{BAAS}%
\def\bac{Bull. astr. Inst. Czechosl.}%
\def\caa{Chinese Astron. Astrophys.}%
\def\cjaa{Chinese J. Astron. Astrophys.}%
\def\icarus{Icarus}%
\def\jcap{J. Cosmology Astropart. Phys.}%
\def\jrasc{JRASC}%
\def\mnras{MNRAS}%
\def\memras{MmRAS}%
\def\na{New A}%
\def\nar{New A Rev.}%
\def\pasa{PASA}%
\def\pra{Phys.~Rev.~A}%
\def\prb{Phys.~Rev.~B}%
\def\prc{Phys.~Rev.~C}%
\def\prd{Phys.~Rev.~D}%
\def\pre{Phys.~Rev.~E}%
\def\prl{Phys.~Rev.~Lett.}%
\def\pasp{PASP}%
\def\pasj{PASJ}%
\def\qjras{QJRAS}%
\def\rmxaa{Rev. Mexicana Astron. Astrofis.}%
\def\skytel{S\&T}%
\def\solphys{Sol.~Phys.}%
\def\sovast{Soviet~Ast.}%
\def\ssr{Space~Sci.~Rev.}%
\def\zap{ZAp}%
\def\nat{Nature}%
\def\iaucirc{IAU~Circ.}%
\def\aplett{Astrophys.~Lett.}%
\def\apspr{Astrophys.~Space~Phys.~Res.}%
\def\bain{Bull.~Astron.~Inst.~Netherlands}%
\def\fcp{Fund.~Cosmic~Phys.}%
\def\gca{Geochim.~Cosmochim.~Acta}%
\def\grl{Geophys.~Res.~Lett.}%
\def\jcp{J.~Chem.~Phys.}%
\def\jgr{J.~Geophys.~Res.}%
\def\jqsrt{J.~Quant.~Spec.~Radiat.~Transf.}%
\def\memsai{Mem.~Soc.~Astron.~Italiana}%
\def\nphysa{Nucl.~Phys.~A}%
\def\physrep{Phys.~Rep.}%
\def\physscr{Phys.~Scr}%
\def\planss{Planet.~Space~Sci.}%
\def\procspie{Proc.~SPIE}%
\let\astap=\aap
\let\apjlett=\apjl
\let\apjsupp=\apjs
\let\applopt=\ao
\bibliography{gc}

\end{document}